# Deep Learning–Based Characterization of Detonation-Cell Size Distributions in Soot-Foil Records

Mingyang Bu[1], Robson A. Schneider[2], Karl P. Chatelain[3,*], Mhedine Alicherif[3,*], Yingchen Shi[1], Andrés Z. Mendiburu[2], Deanna A. Lacoste[3], Bing Wang[4,*]

[1]School of Aerospace Engineering, Tsinghua University, Beijing 100084, China.

[2]Mechanical Engineering Department, Federal University of Rio Grande do Sul (UFRGS), Rua Sarmento Leite 425, Porto Alegre, Brazil.

[3]Physical Science and Engineering Division, Mechanical Engineering Program, King Abdullah University of Science and Technology (KAUST), Thuwal, 23955-6900, Saudi Arabia.

[4]Institute for Aero Engine, Tsinghua University, Beijing 100084, China.

[*]Corresponding author: alicherif.mhedine.f2@f.mail.nagoya-u.ac.jp, karl.chatelain@kaust.edu.sa and wbing@tsinghua.edu.cn.

**Abstract:** The geometric size and regularity of detonation cells are key physical parameters for characterizing the chemical kinetic sensitivity and hydrodynamic instability of detonation waves. Traditional manual measurement of soot foils is time-consuming and prone to subjective errors, whereas existing computer vision and semantic segmentation techniques often exhibit poor generalization when processing real experimental images characterized by high noise, blurred boundaries, and severe overlapping. To address this bottleneck, this paper proposes a novel method for the automated recognition and high-order feature extraction of detonation cells based on deep learning instance segmentation (Mask R-CNN). By constructing a custom heterogeneous dataset (comprising both numerical simulations and physical experiments) and integrating transfer learning strategies, the proposed model achieves accurate recognition and pixel-level mask prediction of individual cells within highly noisy and complex flow fields. The results indicate that the model achieves high pixel-level agreement in numerical simulation benchmark validations and exhibits strong robustness against noise when processing complex real-world experimental soot foils. The predicted average cell sizes agree well with manually measured averages, yielding relative errors of less than 2% and 3.5% for regular and irregular conditions, respectively. Sensitivity ablation experiments confirm the model's excellent scale adaptability while revealing the prior distribution shift and relative scale mechanisms of the algorithm when processing ultra-high-density images. Based on these findings, a standardized preprocessing paradigm for appropriate image patching is established. Overcoming the traditional limitation of extracting only global average sizes, this model achieves the automated tracking of the transient spatial evolution of cell sizes along the propagation direction. Furthermore, it quantitatively extracts high-order regularity features, such as the irregularity index (RI) and the standard deviation of cell deflection angles ($\sigma_\theta$), with the extracted data trends demonstrating consistency with theoretical expectations. The proposed method not only enhances the efficiency and objectivity of the statistical analysis of cell geometric features but also provides an efficient data extraction tool for the in-depth analysis of experimental and numerical soot foils.

## Nomenclature

| | |
|---|---|
| λ | Detonation cell size (characteristic width) |
| PDE | Pulse Detonation Engine |
| RDE | Rotating Detonation Engine |
| CFD | Computational Fluid Dynamics |
| CNN | Convolutional Neural Network |
| Mask R-CNN | Mask Region-based Convolutional Neural Network |
| RPN | Region Proposal Network |
| RoI | Region of Interest |
| FPN | Feature Pyramid Network |
| PAFPN | Path Aggregation Feature Pyramid Network |
| FCN | Fully Convolutional Network |
| SGD | Stochastic Gradient Descent |
| BN | Batch Normalization |
| COCO | Common Objects in Context (dataset format) |

# 1. Introduction

The three-dimensional cellular structure of detonation originates from the coupled interaction between shock waves and chemical reactions. The intersection of transverse waves with the leading shock front forms "triple points" characterized by localized high temperatures and elevated pressures. As the detonation wave propagates forward, these triple points undergo continuous periodic transverse collisions and reflections along the wave front. When recorded on a wall (or surface), the interwoven diamond-shaped trajectories they trace out constitute the macroscopic cellular structure[1–3]. Currently, methods for capturing detonation cellular structures primarily fall into two major categories: experimental observations and numerical simulations. In terms of experimental measurement, the most classical and intuitive approach is the soot track method (soot foil technique). By lining the inner wall of a detonation tube with soot-coated metal foils, the high-temperature and high-pressure triple points on the wave front scrape off the soot, thereby directly recording the diamond-shaped physical trajectories[4–7]. On the other hand, high-fidelity computational fluid dynamics (CFD) simulations have emerged as a crucial tool. By solving hydrodynamic equations coupled with complex chemical kinetics, numerical soot foils can be obtained from the maximum pressure (and/or temperature) fields recorded during the simulation, from which the triple-point trajectories and the cellular structure evolution can be reconstructed[8–12].

Cell size (typically denoted by λ) is of vital importance for the diagnostics of detonation combustion flow fields. First, it serves as a key physical parameter characterizing the detonation sensitivity of combustible mixtures; a smaller cell size indicates highly reactive chemical kinetics, easier initiation, and enhanced resistance to perturbations[13,14]. Secondly, this dimension directly determines macroscopic critical dynamic parameters (e.g., critical initiation energy and tube diameter)[15], enabling accurate predictions of initiation and failure behaviors via geometric scaling relationships. Furthermore, in engineering-oriented aerospace propulsion, such as pulse and rotating detonation engines (PDEs/RDEs), the characteristic geometric dimensions of the combustor must strictly match the cell size to guarantee stable self-sustained propagation[16–18].

Given its critical importance, extensive research has sought to automate cell size measurement to overcome the inherent subjectivity and errors of manual methods[1]. Early automated approaches primarily relied on digital image and signal processing techniques—such as power spectral density combined with edge detection[19,20], autocorrelation function analysis[21], and two-dimensional Fourier spectral analysis of optical diagnostic images[22] —alongside indirect estimation methods[23]. With the rapid advancement of machine learning (ML), various predictive models, including support vector regression (SVR) and artificial neural networks (ANNs), have been developed[24–27]. However, rather than directly analyzing complex experimental soot foil images, these early ML models functioned primarily as parametric regression tools, relying on macroscopic initial conditions (e.g., initial pressure, equivalence ratio, and fuel volume fraction) to indirectly predict the detonation cell size.

To achieve the efficient and high-precision automated extraction of cell geometric features from real experimental images, driven by the rapid advancement of modern computer vision and deep learning-based image processing technologies[28], the research focus has gradually shifted toward direct intelligent image segmentation and feature recognition. Jalontzki et al.[29] proposed an automated analysis method for detonation cells based purely on computer vision. Through a four-step pipeline comprising image preprocessing, cell contour detection, automated parameter optimization, and statistical analysis, this method accomplishes cell recognition, length and width measurements, and size distribution statistics for both numerical and experimental soot foil images without requiring training data. Drawing inspiration from Cellpose—a segmentation model used in cell biology—Sharma et al.[30] proposed a deep learning-based cell segmentation algorithm that operates without the need for training datasets or manual annotations. By integrating image preprocessing, a modified

U-Net model (semantic segmentation), and a feature extraction workflow, this approach successfully segments detonation cells in both experimental and numerical soot foil images.

Despite the progress achieved by the aforementioned image analysis methods based on traditional computer vision or preliminary deep learning, their performance may still be limited when handling real experimental images characterized by high noise, severely blurred boundaries, or severe cell overlapping, where challenges such as reduced generalization or the need for extensive post-processing can arise. Consequently, achieving the accurate segmentation and statistical analysis of individual cells remains highly challenging. Compared to traditional semantic segmentation, which is restricted to pixel-wise category classification[31], instance segmentation technology possesses the capability to accurately distinguish and isolate densely packed and overlapping individual instances of the same class[32]. When processing complex and highly noisy experimental soot foil images, instance segmentation not only effectively overcomes the misidentification of connected components caused by irregular and complex boundaries but also enables the automated and high-precision statistical distribution analysis of geometric features—such as length, width, and area—for every independent cell. To address the limitations of current research, this paper proposes a novel method for detonation cell recognition based on deep learning instance segmentation. By independently constructing a custom training dataset of detonation cells and employing a transfer learning strategy, the segmentation network is trained on the basis of generalized pre-trained weights. This approach successfully achieves the recognition and pixel-level mask prediction of individual cells within highly noisy and complex cell images. Furthermore, it statistically extracts the key geometric features of each independent cell, thereby providing an efficient, objective, and intelligent tool for the quantitative analysis of the macroscopic characteristic dimensions in detonation combustion.

# 2. Cell identification model and method

This chapter primarily introduces the deep learning model architecture and its training scheme designed for cell image instance segmentation. Adapting to the specific characteristics of cell images, this study constructs a targeted design and parameter optimization based on the Mask R-CNN (Mask Region-based Convolutional Neural Network) two-stage object detection and segmentation framework[33]. This chapter details the network architecture design, data augmentation mechanisms, and the training strategies combined with hyperparameter configurations.

## 2.1 Network Architecture Design

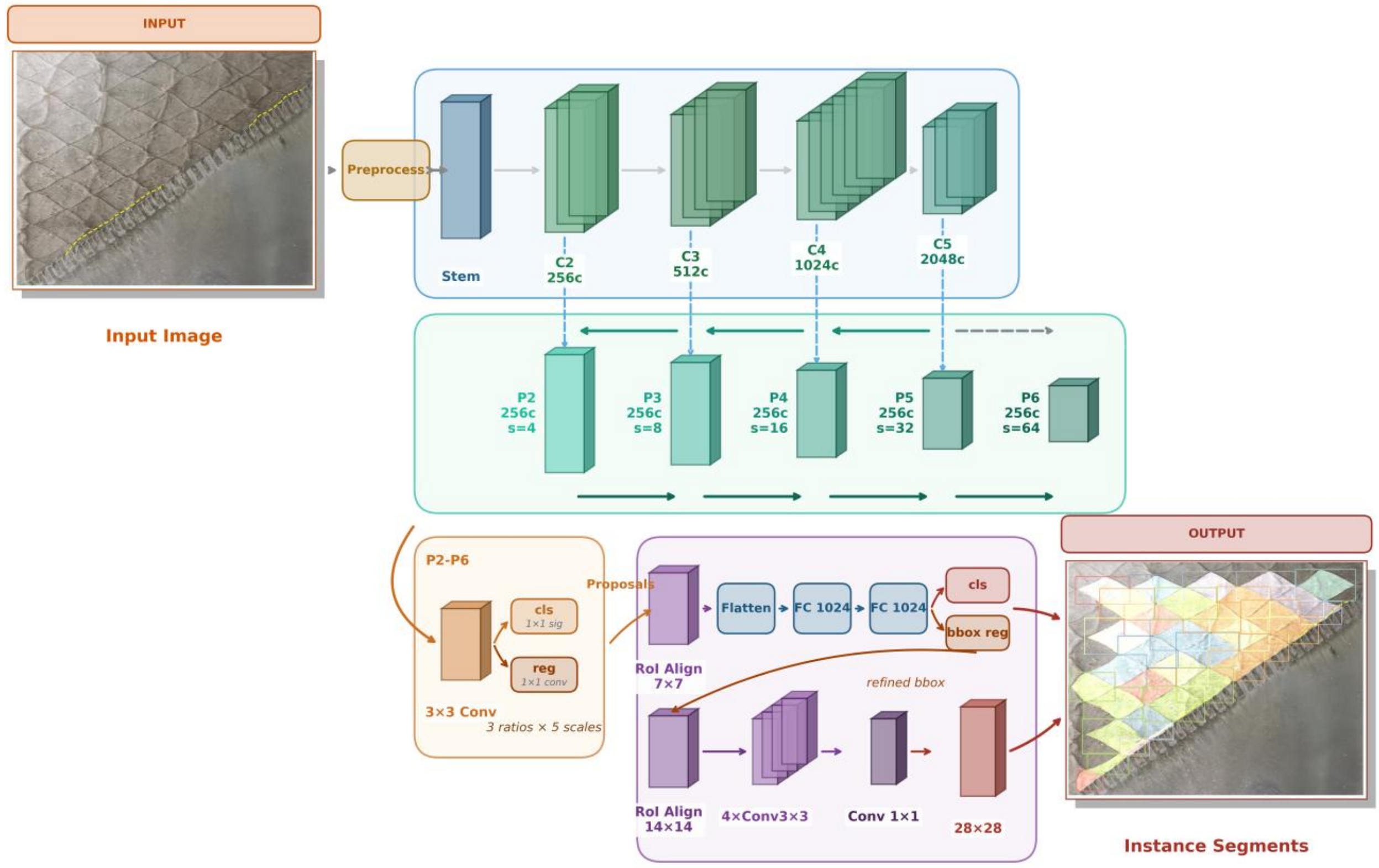

Figure 1 Network architecture

The instance segmentation model employed in this study utilizes Mask R-CNN as the overarching framework, which mainly consists of four core modules: the Backbone, the Neck, the Region Proposal Network (RPN), and the Region of Interest (RoI) Head. For the single-class segmentation task (i.e., Cell), the network structure is meticulously designed as follows:

### A. Backbone Network

The deep residual network ResNet-50 is adopted as the foundational feature extractor. To accelerate model convergence and enhance generalization capability, the backbone is initialized with weights pre-trained on the ImageNet dataset. The network comprises 4 stages, outputting feature maps at four different scales to capture the multiscale morphological information of cells. All Batch Normalization (BN) layers are configured to be trainable (unfrozen).

### B. Neck (Multiscale Feature Fusion)

A Path Aggregation Feature Pyramid Network (PAFPN) is utilized as the neck architecture. It receives feature maps from the backbone with channel dimensions of

[256,512,1024,2048] and uniformly reduces them to 256 channels. PAFPN enhances the fusion of low-level spatial information and high-level semantic information through both top-down and bottom-up pathways, thereby significantly improving the representation capability for cells of varying sizes and morphologies.

**C. Region Proposal Network (RPN)**

The RPN is responsible for generating potential candidate regions (Proposals) containing cells on the feature maps. The model is equipped with a multiscale Anchor Generator, where the base scales are set to [1,2,4,8,16], and the aspect ratios are designed as [0.5,1.0,2.0], corresponding to feature strides of [4,8,16,32,64]. This configuration is highly suitable for the morphological distribution of both elongated and circular cells.

**D. Region of Interest Network (RoI Head)**

The feature extraction for candidate regions employs the RoIAlign algorithm for pixel-level alignment, which resolves the quantization error introduced by traditional RoIPooling and is crucial for high-precision mask generation. BBox and classification head adopts the Shared2FCBBoxHead. After the feature maps are pooled to 7×7, they pass through fully connected layers to output classification and bounding box regression results. Mask head employs a Fully Convolutional Network (FCN) mask head (FCNMaskHead). The feature maps are pooled to 14×14 and undergo upsampling and feature decoding through 4 consecutive convolutional layers, ultimately outputting a single-class pixel-level segmentation mask.

## 2.2 Training and Transfer Learning Strategies

Given that detonation cell images typically face challenges such as the high difficulty of soot foil experiments and limited annotated sample sizes, training a deep neural network from scratch is highly susceptible to overfitting. To enhance the model's robustness and accelerate convergence, this study adopts a two-stage synergistic transfer learning strategy[34]. Initially, the ResNet-50 backbone is initialized with universal weights pre-trained on a large-scale natural image dataset (ImageNet), endowing the network with robust baseline capabilities for extracting fundamental visual features like edges and textures right from the start. Subsequently, for holistic network optimization, the model avoids random initialization by loading optimal weights obtained from a preliminary experimental stage. This continuous fine-tuning strategy within the target physical domain drastically reduces the "cold start" period and guides the network to converge more precisely towards the local optimum for complex detonation cell structures.

During the data preprocessing and augmentation stages, strictly targeted designs are implemented based on the unique characteristics of detonation cellular trajectory images. All input images are consistently resampled to a resolution of 1024×1024 pixels to optimally preserve the fine microscopic trajectory details left by transverse shock wave collisions. To further increase data diversity and improve model generalization, a subset of the training data incorporates a rigorous augmentation pipeline, which includes combined random flipping in both horizontal and vertical directions, as well as PhotoMetric Distortion. These enhancement methods effectively simulate the natural morphological distortions of detonation cells, variations in triple-point trajectory distributions, and visual contrast deviations caused by uneven illumination or varying soot coating thicknesses typical in real-world experiments (e.g., soot foil records).

The parameter optimization of the model is driven by a carefully designed joint loss function, with balanced weights set to 1.0 for all individual components. Specifically, the classification tasks of both the Region Proposal Network (RPN) and the Region of Interest (RoI) employ Cross-Entropy Loss, with the RPN dynamically utilizing a weighted Sigmoid function. For the bounding box coordinate regression representing the cell scales, Smooth L1 Loss is applied to enhance the network's robustness against irregular local cell localizations. Finally, the detailed mask generation module also utilizes Cross-Entropy Loss to rigorously supervise pixel-level classification, thereby ensuring the high fidelity and structural accuracy of the final cellular grid boundaries.

The batch size limitations of the framework and the optimization algorithm directly determine the network's optimization trajectory. This experiment relies on the Stochastic Gradient Descent (SGD) optimizer for parameter updates. To prevent trapping in local optima while maintaining training stability, we integrated a hybrid learning rate scheduling strategy combining Linear Warmup and Cosine Annealing. This configuration provides a robust learning foundation for high-density and multiscale object segmentation in microscopic cell images.

## 2.3 Dataset Introduction

The dataset employed in this study is constructed from heterogeneous multi-source data, encompassing not only actual detonation cellular images acquired through physical experiments (such as classical soot foil recording) but also flow field evolution results generated via high-fidelity numerical simulations[35–37]. This combination of experimental and computational data effectively enriches the sample space and improves the model's generalization capability across varying and complex working conditions. To comply with the training protocols of the Mask R-CNN instance segmentation model, the original images from all sources and their corresponding manual annotations are uniformly organized into the standard COCO (Common Objects in Context) dataset format.

Structurally, the dataset is rigorously partitioned into a training set (80%) and a validation set (20%) at a ratio of 8:2. This strict separation ensures sufficient data for model learning while facilitating unbiased hyperparameter updates and objective performance evaluation. During the data feeding and preprocessing phase, the model synchronously decodes images and JSON-formatted annotation files via a predefined loading pipeline. To unify the computational dimensions while optimally preserving the fine microscopic structures inherent in both soot foil records and simulated flow fields, all input images are resized and padded to a standardized spatial resolution of 1024×1024 pixels while strictly maintaining their original aspect ratios. Furthermore, channel-wise (RGB) normalization is applied to pixel values using predefined channel means and standard deviations. This operation effectively mitigates domain shifts caused by either experimental illumination variations or simulation pseudo-color mappings, ensuring numerical stability during deep feature extraction.

Upon the completion of training, a validation set encompassing a broader range of operating conditions was utilized for the final evaluation of the model's performance [35,38–40]. This was conducted to fully demonstrate the robustness and generalizability of the model, particularly when processing experimental soot foil images characterized by high noise levels and complex cell boundaries.

# 3. Results and Discussion

## 3.1 Identification effect display

To quantitatively and objectively evaluate the performance of the proposed instance segmentation model, benchmark validation is first conducted using a cell dataset equipped with absolute ground truth. Specifically, numerically simulated cells—encompassing both regular and irregular configurations—are utilized to verify the model's recognition performance, as depicted in Figure 2 and Figure 3. During the model prediction and post-processing stage (as illustrated in Figure 3(b)), incomplete cells situated at the outer margins of the image are excluded by the algorithm due to boundary truncation effects. This design is crucial because truncated cells would artificially introduce undersized geometric dimensions, thereby compromising the statistical authenticity of the size probability density distribution. It is evident that for regular simulated cells, the model's recognition performance achieves a highly satisfactory pixel-level agreement. Visualization results indicate that the predicted cell masks exhibit sharp edges and clear boundary delineations, with zero instance omission (i.e., an exceptionally high recall rate). This not only verifies the model's capability to capture the features of ideal detonation wave trajectories but also demonstrates the high efficiency of the RoIAlign module within the network for pixel alignment. More challenging are the irregular simulated cells presented in Figure 3. These cells simulate the instability and complex topological structures of transverse wave collisions on actual detonation wave fronts. As illustrated in the $\lambda$ distribution histogram in Figure 3(c), although minor discrepancies exist between the model's predicted probability distribution and the ground truth within certain size intervals—primarily attributable to the intrinsic ambiguity in defining pixel-level masks caused by the highly distorted boundaries of irregular cells—the overall distribution trend remains highly consistent. More importantly, the average cell size predicted by the model (37.61px) nearly coincides with the ground truth mean (37.64px), yielding a negligible error.

Simultaneously, Figure 3 (b) further confirms that even under conditions of distorted and overlapping cell morphologies, the model—benefiting from the robustness of the Region Proposal Network (RPN)—still achieves zero-omission individual segmentation. These results fully demonstrate that when processing complex and irregular detonation flowfield structures, the proposed deep learning model not only possesses robust feature generalization capabilities but also provides high-precision macroscopic statistical dimensions. This lays a solid foundation for the subsequent analysis of highly noisy, real-world experimental soot foil images.

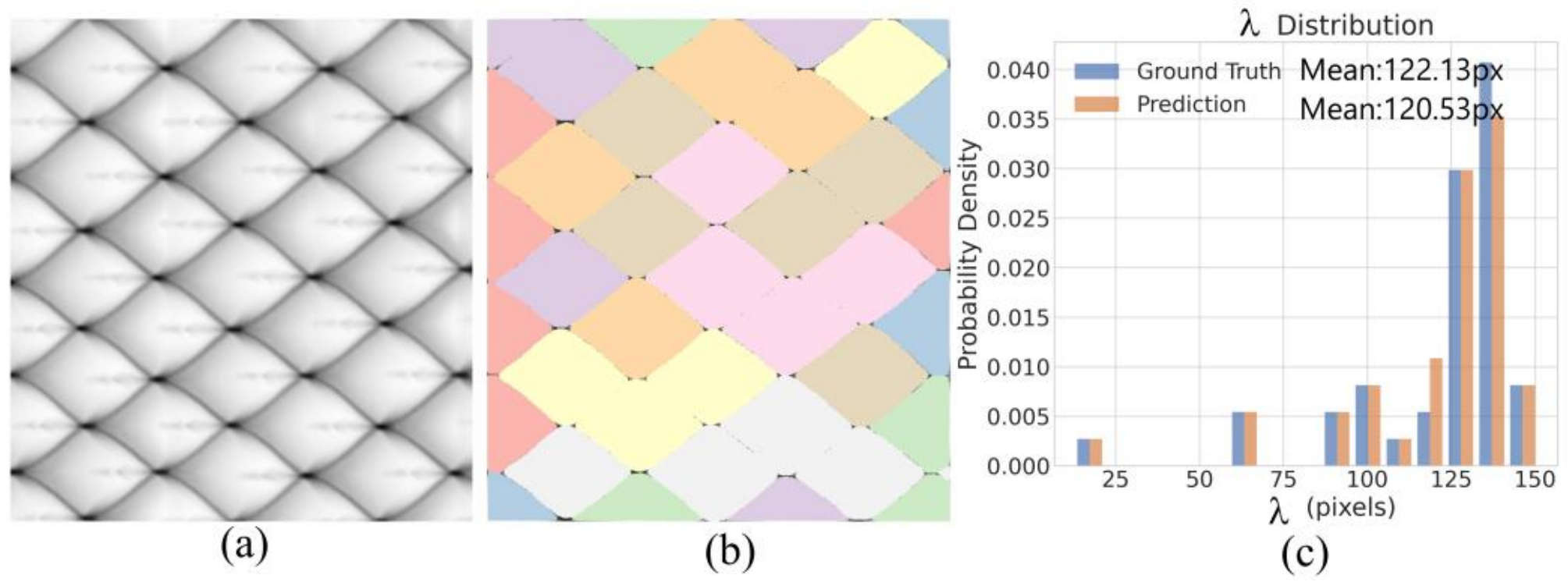

Figure 2 Recognition results of regular simulated cells: (a) original image; (b) model prediction results; (c) cell λ distribution.

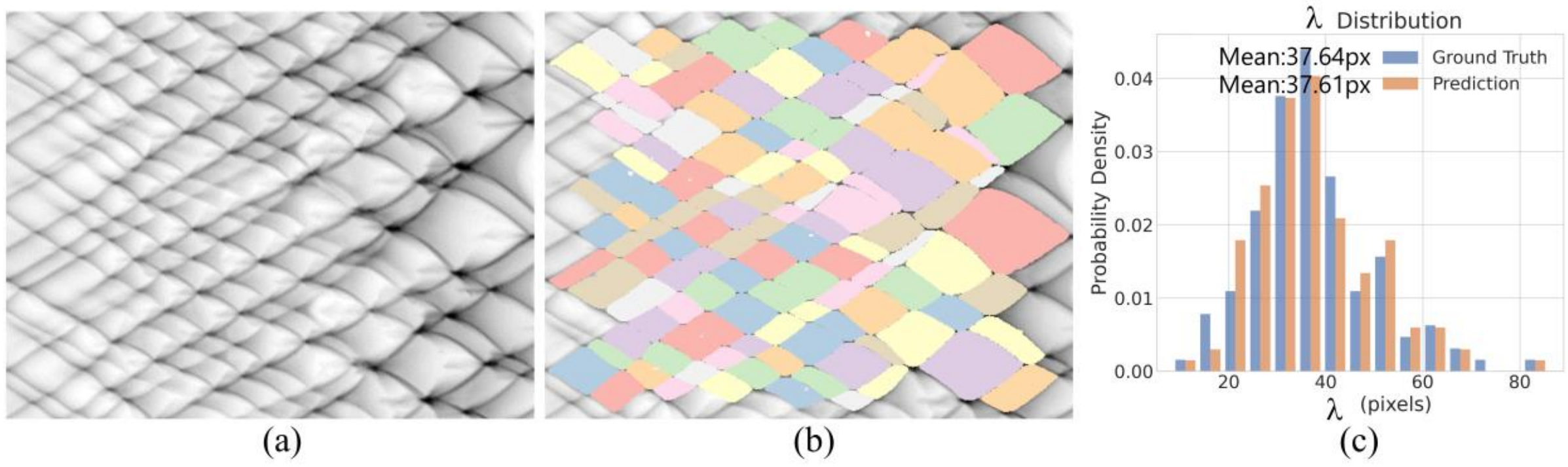

Figure 3 Recognition results of irregular simulated cells: (a) original image; (b) model prediction results; (c) cell λ distribution.

Having validated the model's recognition capability on ideal numerically simulated data, this study further applies it to the more challenging task of feature extraction from real-world experimental soot foil images. As illustrated in the visualization results in Figure 4(b) and Figure 6(b), the original elongated soot foil images are not fed into the network in their entirety. Instead, they are uniformly pre-partitioned into a 2×3 grid of sub-patches for independent inference. This preprocessing strategy is adopted to mitigate the model's sensitivity limitations regarding extremely high object densities within a single image. When the full image is input directly, the inherent compression of feature map resolutions causes densely packed, small-sized cells to easily lose their spatial features during the convolutional and pooling processes. By conducting patch-based inference, the model is guaranteed to operate under optimal resolutions and receptive fields, thereby maximizing the precision and fineness of the predicted segmentation masks. The sensitivity mechanisms governing the relationship between input resolution and the model's recognition performance will be thoroughly investigated and discussed through detailed ablation experiments in Section 3.2.

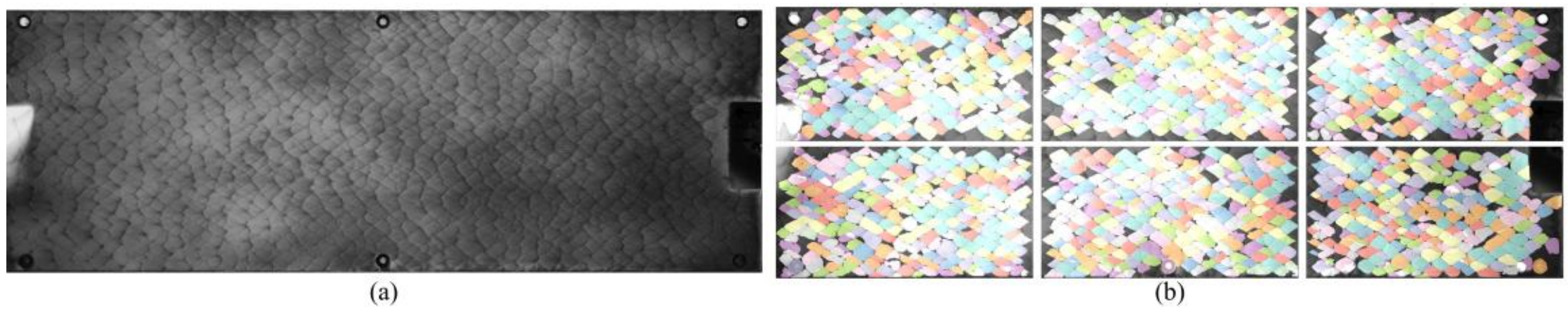

Figure 4 Visualization of recognition results for irregular experimental cells: (a) original image; (b) model prediction results.

Regarding the irregular experimental cells(Figure 4), the images suffer from severe quality degradation, including high background noise, low local contrast caused by uneven soot deposition, and structural complexity due to the intertwining of non-ideal triple-point trajectories. Despite these substantial perturbations, as observed in Figure 4(b), the model still successfully recognizes the vast majority of overlapping cell instances, demonstrating robust feature extraction capabilities against noise. According to the quantitative comparison in Figure 5, the manually annotated reference comprises 1,485 cells with an average size λ of 139.1px, whereas the model predicts and recognizes 1,398 cells with an average size of 144.0px. The probability density distributions of both sets(Figure 5a) exhibit a high degree of overlap, yielding a relative error in the mean value of only 3.4%. Notably, the number of cells predicted by the model (n=1,398) is slightly lower than the manual reference count, accompanied by a minor shift in the mean value. These discrepancies primarily stem from two objective factors: First, the original image is partitioned into a 2 × 3 grid of sub-images for independent recognition before the results are aggregated. Cells crossing the partition lines are inevitably split into incomplete fragments, which are strictly filtered out by the post-processing algorithm to ensure the accuracy of geometric measurements. Since manual annotation is performed directly on the intact original image and is free from such truncation, this discrepancy inherently leads to a lower cell count reported by the model. Second, in actual soot foil experiments, the triple-point tracks in certain regions suffer from severe physical abrasion or extreme contrast degradation. During manual annotation, researchers often leverage prior experience to subjectively connect and infer these blurred trajectories (which conversely results in the appearance of numerous highly dispersed outliers in the size distribution within the reference box plot in Figure 5b). In contrast, the instance segmentation model strictly relies on the confidence thresholds of the feature maps. For severely blurred regions lacking sufficient visual feature support, the model is unable to generate high-quality Region Proposals (RPN), thereby exhibiting a certain degree of omission (missed detections).

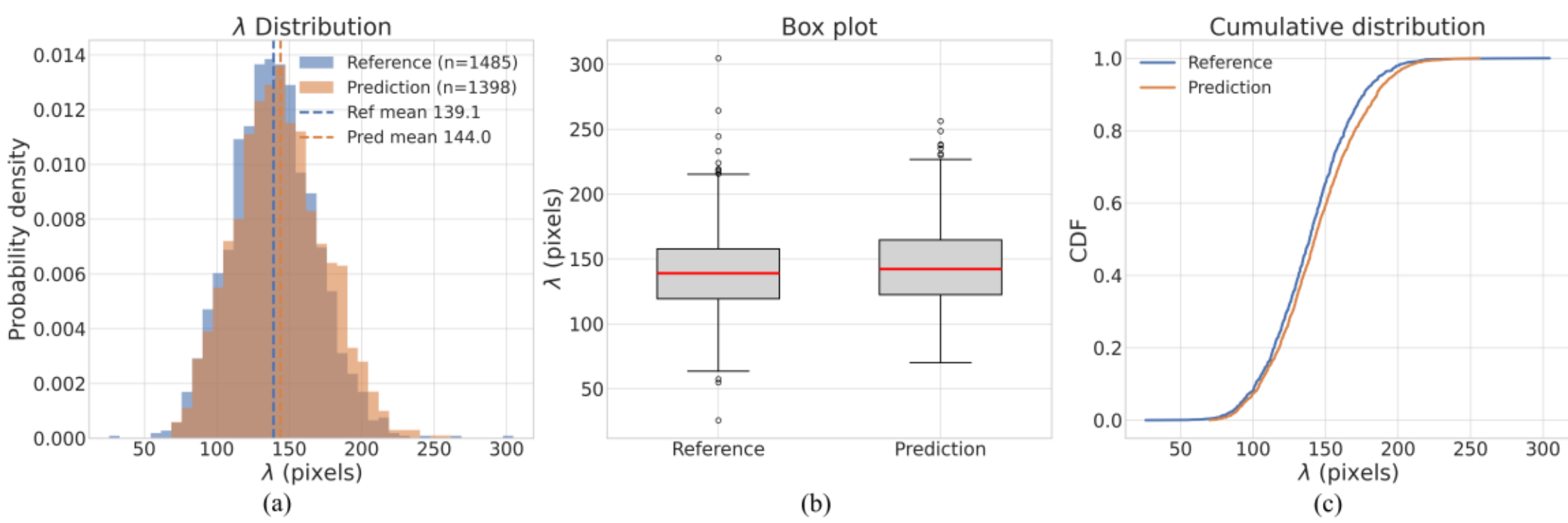

Figure 5 Quantitative comparison of irregular experimental cell recognition: (a) cell λ distribution; (b) box plot; (c) cumulative distribution.

For the experimental cells with relatively regular morphologies( Figure 6), due to the clearer transverse wave trajectories, the model's mask predictions exhibit superior boundary adherence. In terms of quantitative analysis( Figure 7), the average cell size predicted by the model (183.0px) agrees well with the manual reference value (186.6px), yielding a relative error of less than 2%. Similar to the irregular cases, constrained by the boundary truncation effects from image patching and missed detections in local regions with weak features, the total number of cells recognized by the model( n=706) is lower than the manual annotation count for the full image( n=917). However, as can be observed from the Cumulative Distribution Function (CDF) curves in Figure 7(c), this discrepancy in quantity does not negatively impact the overall distribution pattern, with the upward trends of both curves aligning closely. More crucially, observing the box plot in Figure 7(b) reveals that because the model performs extraction strictly based on high-confidence features, its predicted Interquartile Range (IQR) is more concentrated. This effectively avoids the forced fitting of blurred noise into abnormally small or excessively large cells. Furthermore, even in the presence of obvious vertical truncations or edge interference from non-cellular structures in the original experimental images, the model is not misled by these pseudo-boundaries. It remains capable of accurately filtering out the interference and identifying the genuine cell contours, thereby further verifying the outstanding robustness of its feature extraction algorithm.

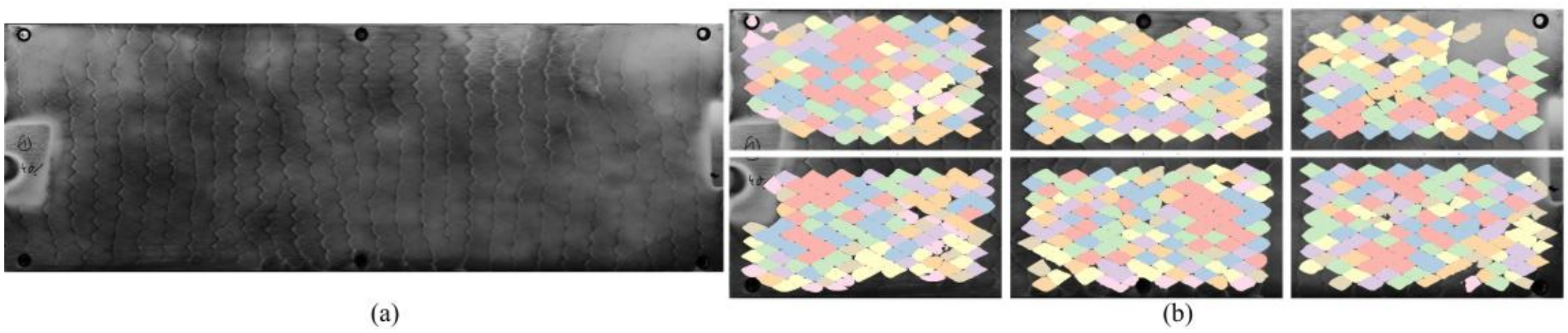

Figure 6 Visualization of recognition results for regular experimental cells: (a) original image; (b) model prediction results.

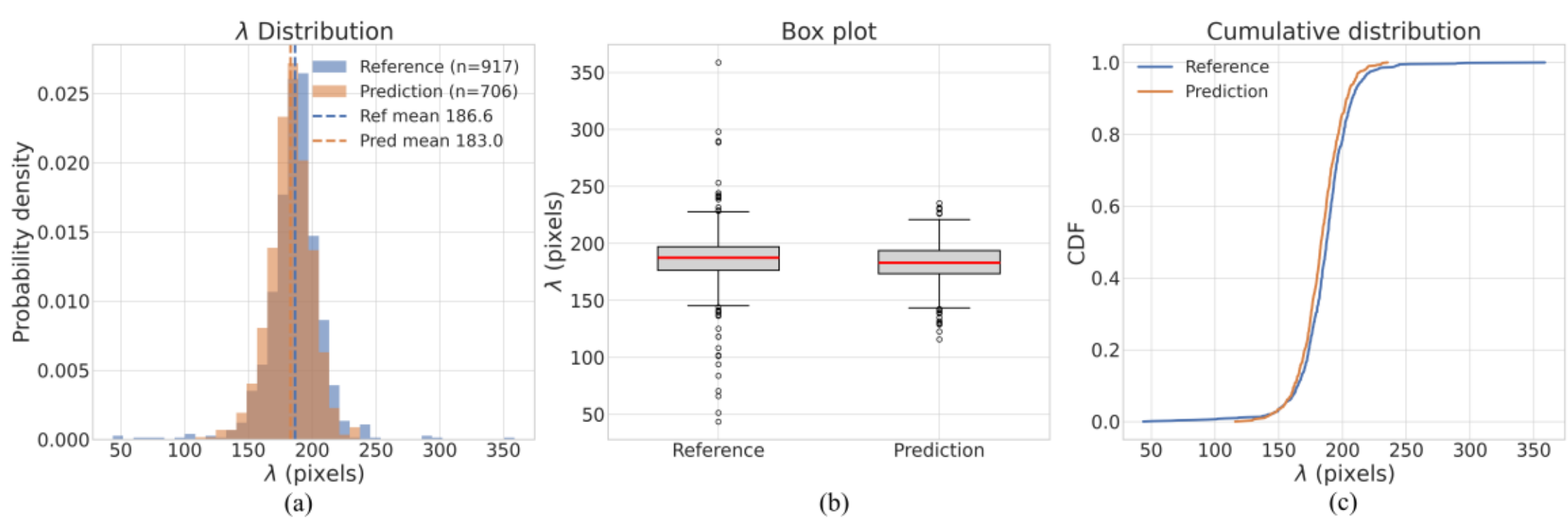

Figure 7 Quantitative comparison of regular experimental cell recognition: (a) cell λ distribution; (b) box plot; (c) cumulative distribution.

## 3.2 Sensitivity analysis of Model

To ensure the accuracy and robustness of the instance segmentation model in practical physical applications, this section conducts sensitivity and ablation analyses on two critical preprocessing parameters that govern the model's predictive performance: input image resolution and per-image object density (image patching strategy).

### 3.2.1 Sensitivity Analysis of Image Resolution

In actual detonation soot foil experiments, constrained by the field of view (FOV) and physical limitations of the imaging equipment, the resolution of the acquired images often varies. To evaluate the model's adaptability to resolution variations, the original image (Figure 4a) was proportionally downsampled to 60% and 30%, respectively. A quantitative comparison of the recognition results was then conducted, with the findings presented in Figure 8.

As evidenced by the quantitative data in Figure 8, the model demonstrates excellent scale invariance and exceptional robustness against noise. Despite the significant compression in image resolution, the predicted probability density distributions and cumulative distribution function (CDF) curves remain highly consistent with their respective reference values. Regarding the mean value predictions, compared to the relative error of approximately 3.5% under the baseline condition (Figure 5), the relative error between the predicted mean (86.6px) and the reference mean (83.5px) at 60% resolution experiences only a marginal increase to roughly 3.7%. Even at a low resolution of 30%, the prediction error is well maintained at around 5% (predicted mean of 43.8px vs. reference mean of 41.7px). Furthermore, the total number of cells predicted by the model at 60% and 30% resolutions is 1,433 and 1,316, respectively. These counts remain of the same order of magnitude as the baseline condition (n=1,398), indicating that no widespread omission (missed detections) occurs as a result of pixel blurring.

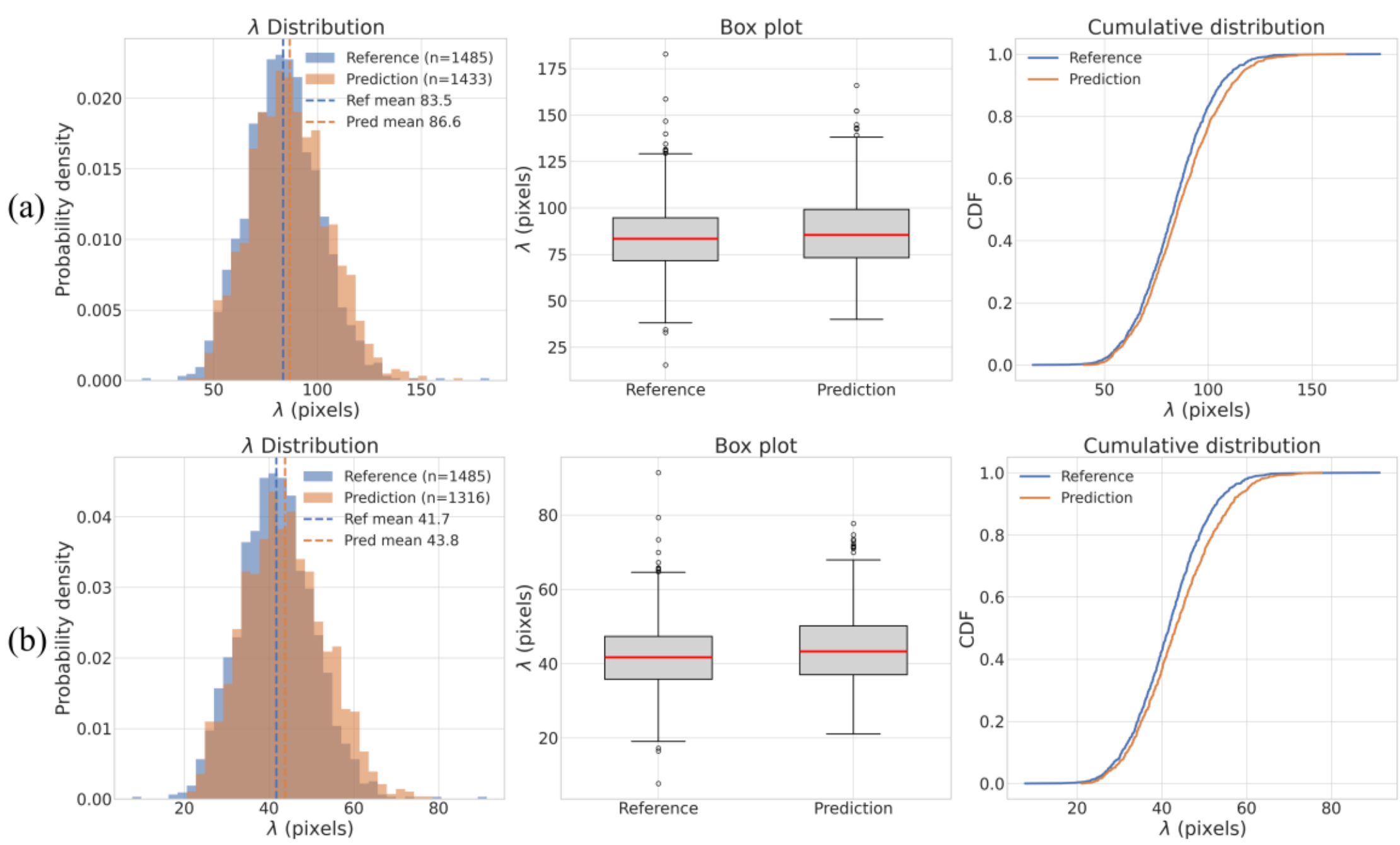

Figure 8 Quantitative comparison of recognition results for the same image (Figure 4a) at different resolutions: (a) 60% resolution; (b) 30% resolution.

The marginal upward shift in the mean value observed in the data aligns with the physical intuition underlying downsampling in Convolutional Neural Networks (CNNs): as the resolution decreases, a minute fraction of cell features—which are inherently small and possess blurred boundaries—are further smoothed out. Driven by confidence thresholding considerations, the model filters out these weak feature patches, thereby statistically resulting in a slight elevation of the mean value. Overall, however, resolution degradation exerts a negligible impact on the general statistical trends of the model. This ablation study demonstrates that the proposed instance segmentation model does not strictly require ideal, ultra-high-resolution inputs. Instead, it can consistently yield high-precision size statistics across a broad dynamic range of resolutions.

### 3.2.2 Sensitivity Analysis of Per-Image Object Density and Patching Strategy

In addition to resolution, the absolute cell count within a single input image constitutes another core factor capable of inducing performance bottlenecks in the model. By comparing direct full-image input(Figure 9) and 1 × 2 partitioned input(Figure 10), and subsequently benchmarking them against the baseline 2 × 3 patching strategy (Figure 5), this section elucidates the necessity of adopting a rational image patching strategy.

As observed in Figure 9, when the entire elongated image containing nearly 1,500 cells is directly fed into the model for inference, severe missed detections occur. In stark contrast to the baseline condition in Figure 5—which recognized 1,398 independent cells—the total predicted count for the direct full-image input plummets to merely 629. The visualization results in Figure 9(a) clearly demonstrate that a substantial number of cells are not extracted by the network, resulting in extensive blank regions between the masks. It is noteworthy that, despite the massive scale of missed detections, the predicted mean cell size (185.4px) significantly surpasses the true mean (139.1px). This occurs because, under highly congested cell conditions, the model tends to prioritize the retention of larger cells with the most prominent contour contrast, while systematically discarding the vast majority of small-to-medium-sized cells. Statistically, this bias severely elevates the mean of the overall distribution. When the image is partitioned into a 1 × 2 grid and subjected to re-inference(Figure 10)，the missed detection issue is significantly alleviated; the recognized count rebounds to 1,256, and the mean value drops back to 153.0px. However, compared to the optimal 2 × 3 state in Figure 5, the omission of certain targets still persists.

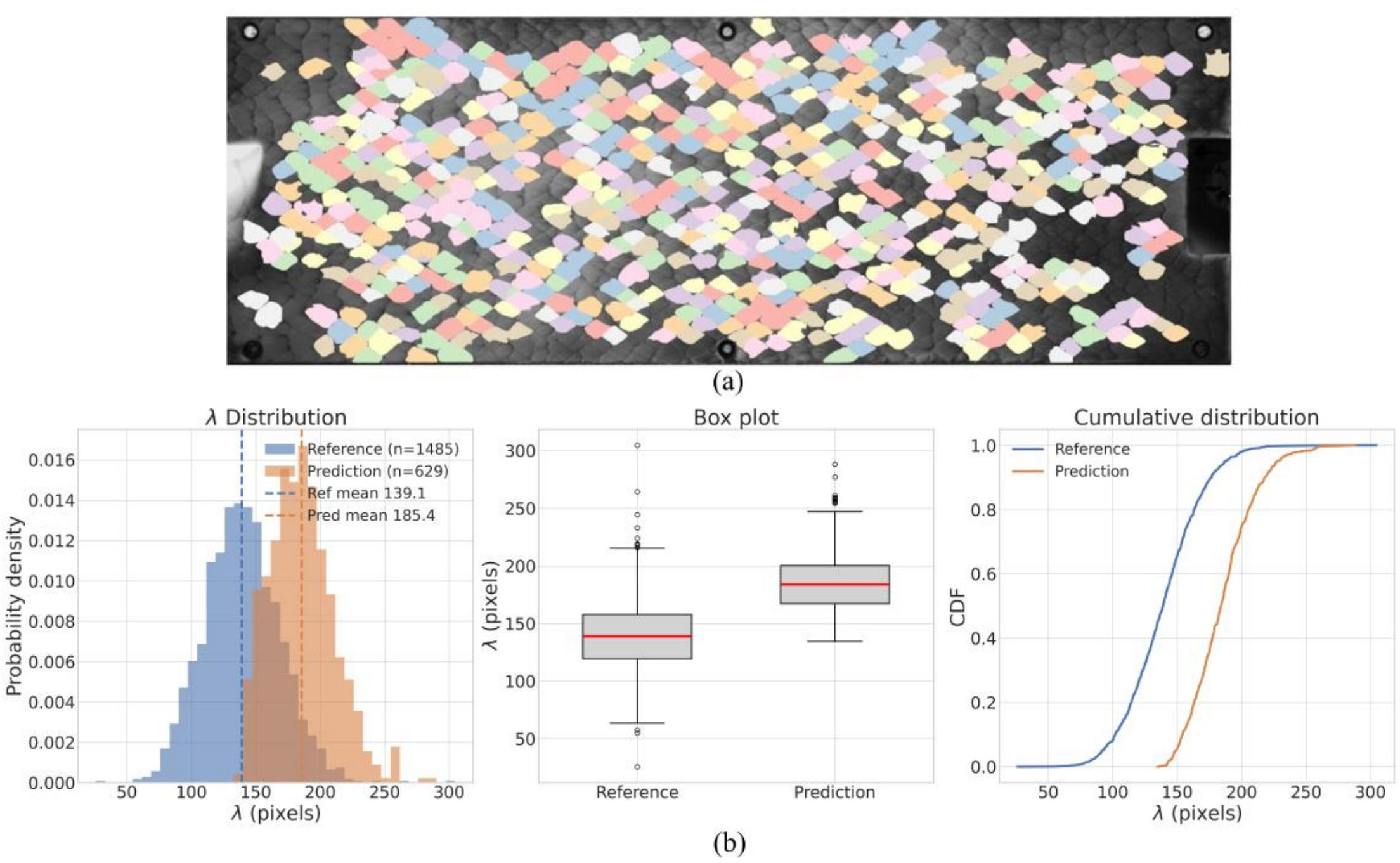

Figure 9 Direct inference on the entire image (Figure 4a): (a) visualization of recognition results; (b) quantitative comparison.

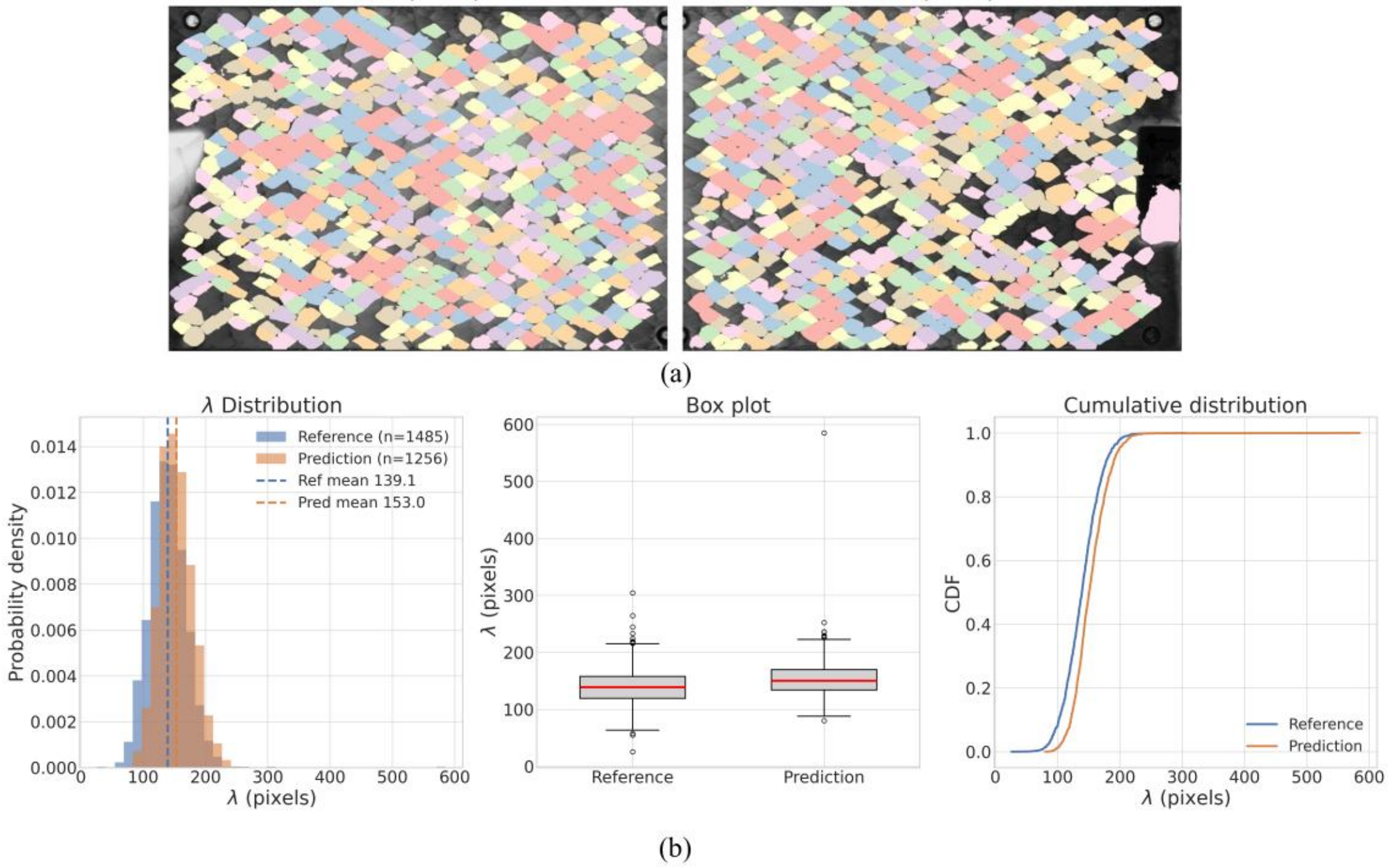

Figure 10 Inference results of the image (Figure 4a) partitioned into a 1 × 2 grid: (a) visualization of recognition results; (b) quantitative comparison.

Having excluded the inherent instance truncation limit of the algorithmic framework (in this study, the maximum number of detectable instances in the underlying network was expanded to 5,000, far exceeding the total count of approximately 1,500 cells present in a single image), the underlying mechanisms responsible for the missed detections during full-image inference can be primarily attributed to feature scale mismatch and prior distribution shift:

Scale Mismatch and Confidence Degradation: During the preprocessing stage, to unify computational dimensions, the model applies standardized resizing to the input images (e.g., resizing to 1024 × 1024). When an entire soot foil image with an extreme aspect ratio is directly input, the relative pixel dimensions of individual cells on the feature map are drastically compressed. This scale contraction causes the minute cell features to severely mismatch with the network's pre-defined anchor scales, leading them to be subsequently smoothed out within the deep convolutional layers. This feature degradation directly results in a drastic decline in the network's predicted confidence scores for these small-to-medium-sized cells. Given that a secondary confidence filtering threshold (e.g., Score > 0.3) and boundary truncation filtering are implemented at the inference application layer, a substantial number of small-to-medium-sized cells—which were initially captured by the network but assigned lower scores—are filtered out during the post-processing stage. Consequently, only a minority of large-scale cells with prominent features are ultimately retained.

Data Distribution Shift: As illustrated by the statistical curves in Figure 11, the number of cells per image within the training set constructed for this study is predominantly concentrated between 50 and 250. Directly inputting the full image (Figure 9) falls far outside the prior density distribution learned by the model during the training phase, thereby inducing a domain shift (Domain Shift). Conversely, by adopting the 2 × 3 patching strategy (Figure 5), each sub-patch contains approximately 200 cells. This falls precisely within the distribution interval where the model is optimally trained and achieves peak performance.

In summary, adopting a rational image patching strategy, such as the 2 × 3 configuration, is not merely an essential measure for maintaining the relative scale of cell features and preventing the anomalous degradation of confidence scores; it is fundamentally the key to ensuring strict alignment between the feature density of the input images and the prior distribution of the training set. This comparative analysis establishes a standardized preprocessing paradigm for the subsequent processing of large-format detonation cell images.

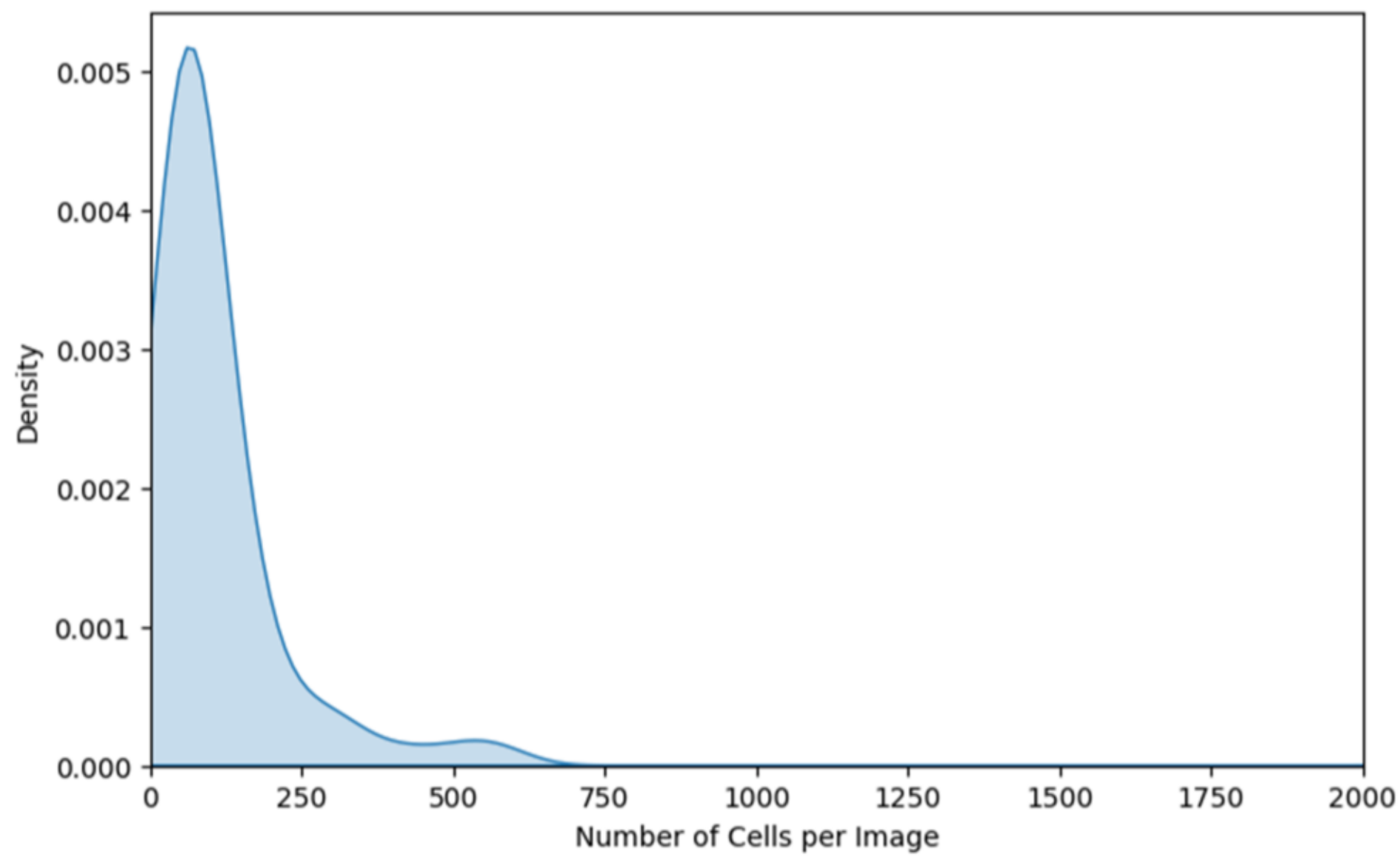

Figure 11 Distribution curve of the number of cells per image in the training set

## 3.3 Advanced extraction capability of cell size information

Existing methods for detonation cell analysis are typically limited to extracting a single global average cell width (λ) across the entire observation region. However, the propagation of a detonation wave is a highly nonlinear transient process, and relying solely on a global average often obscures the localized dynamic evolution and dimensional characteristics of the flow field. Leveraging the capability of the Mask R-CNN instance segmentation model to extract pixel-level coordinates and masks for every independent cell instance, this study successfully achieves the automated data extraction of the spatial evolution of cell sizes, alongside high-order geometric features (such as regularity and deflection angle).

### 3.3.1 Analysis of the Axial Spatial Evolution of Cell Size λ

Investigating the evolution of cell sizes along the propagation direction of the detonation wave (the X-axis) is of significant physical importance for assessing whether the detonation wave has achieved a self-sustained stable state, as well as for analyzing the attenuation or amplification of transverse wave intensity. By utilizing the centroid coordinates and individual widths of each cell output by the model, we can reconstruct a spatial distribution scatter plot illustrating λ versus the propagation distance.

As illustrated in Figure 12, for the ideal regular simulated cells, the λ data points extracted by the model are tightly clustered around the global mean (67.4px). The slope of the linear fit along the propagation direction is minuscule (Slope = +0.0001), and the standard deviations (error bars) within each interval are extremely low. This corroborates the physical fact that the detonation wave is propagating in an absolute steady state under this condition. By contrast, Figure 13 illustrates the spatial evolution patterns of the irregular simulated cells. It is clearly observable that not only does the local dispersion (standard deviation) of the cell sizes increase, but the linear fit also exhibits a distinct positive slope (Slope=+0.0043). This transient spatial trend, captured by the proposed model, suggests that the flow field at this stage may have experienced slight detonation wave decoupling or transverse wave energy dissipation, ultimately resulting in a progressive enlargement of the cell sizes.

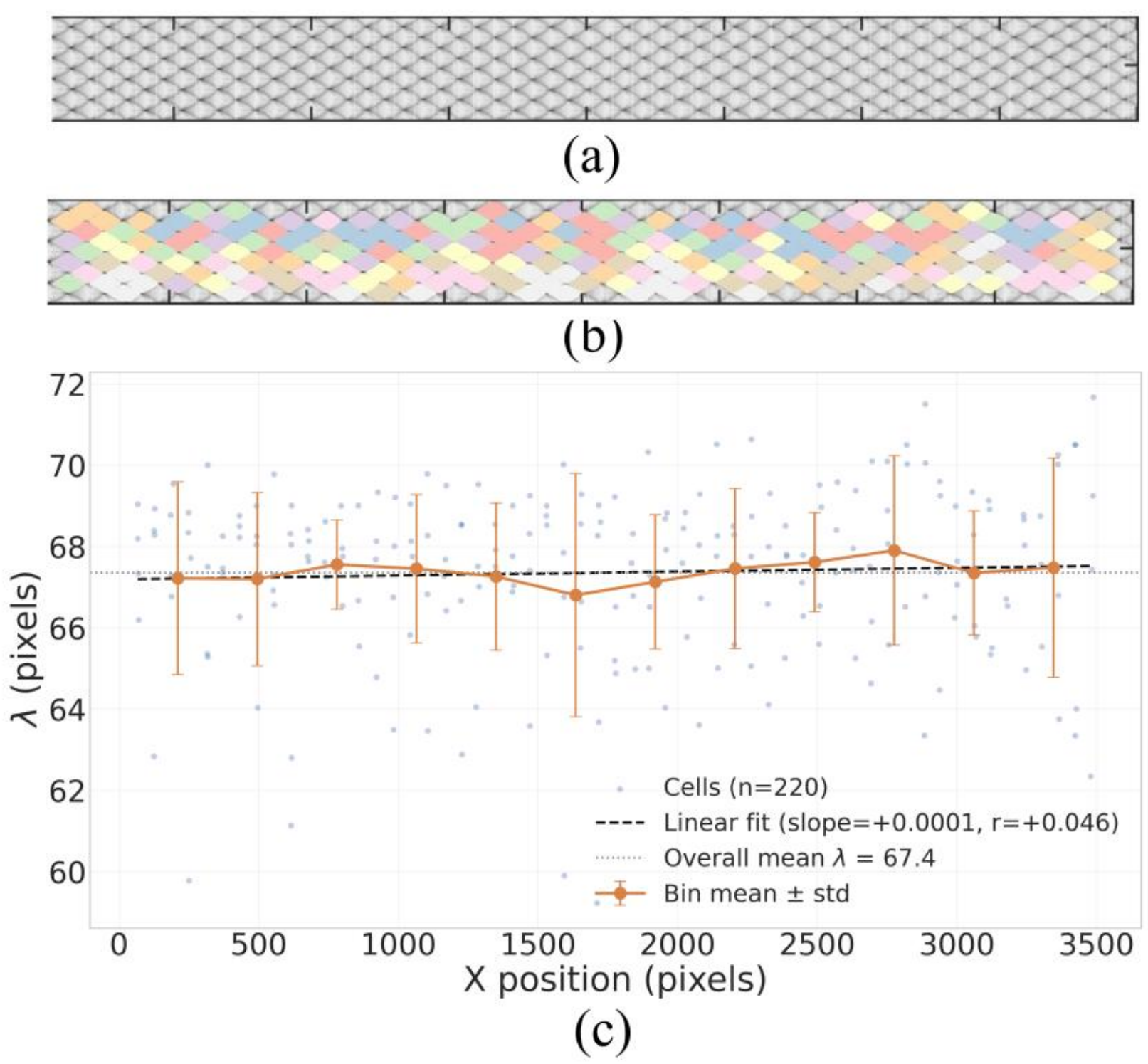

Figure 12 Regular simulated cells: (a) original image; (b) recognition results; (c) variation of λ along the propagation direction of the detonation wave.

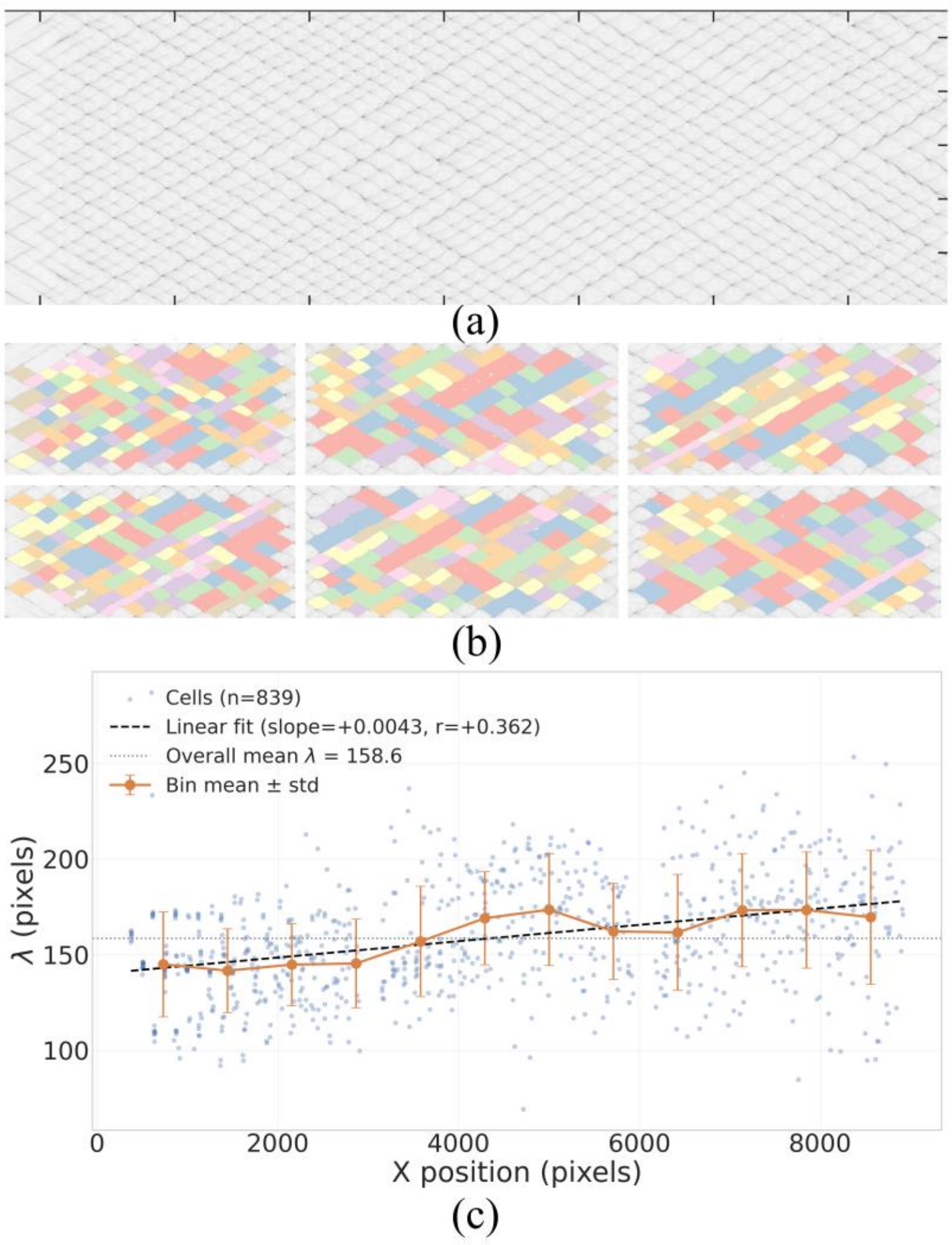

Figure 13 Irregular simulated cells: (a) original image; (b) recognition results; (c) variation of λ along the propagation direction of the detonation wave.

Furthermore, Figure 14 demonstrates the model's extraction capability on the complex, elongated experimental soot foil image. Despite the presence of significant background noise and localized abrasion in the experimental image, the model remains capable of robustly delineating the spatial distribution trajectories of 1,398 cells across the entire field. The fitted slope (Slope = +0.0003) approaches zero, which objectively and quantitatively confirms that, under this experimental condition, the detonation wave has dissipated the overdriven effects characteristic of the initial initiation phase and transitioned into a stable Chapman-Jouguet (CJ) propagation state.

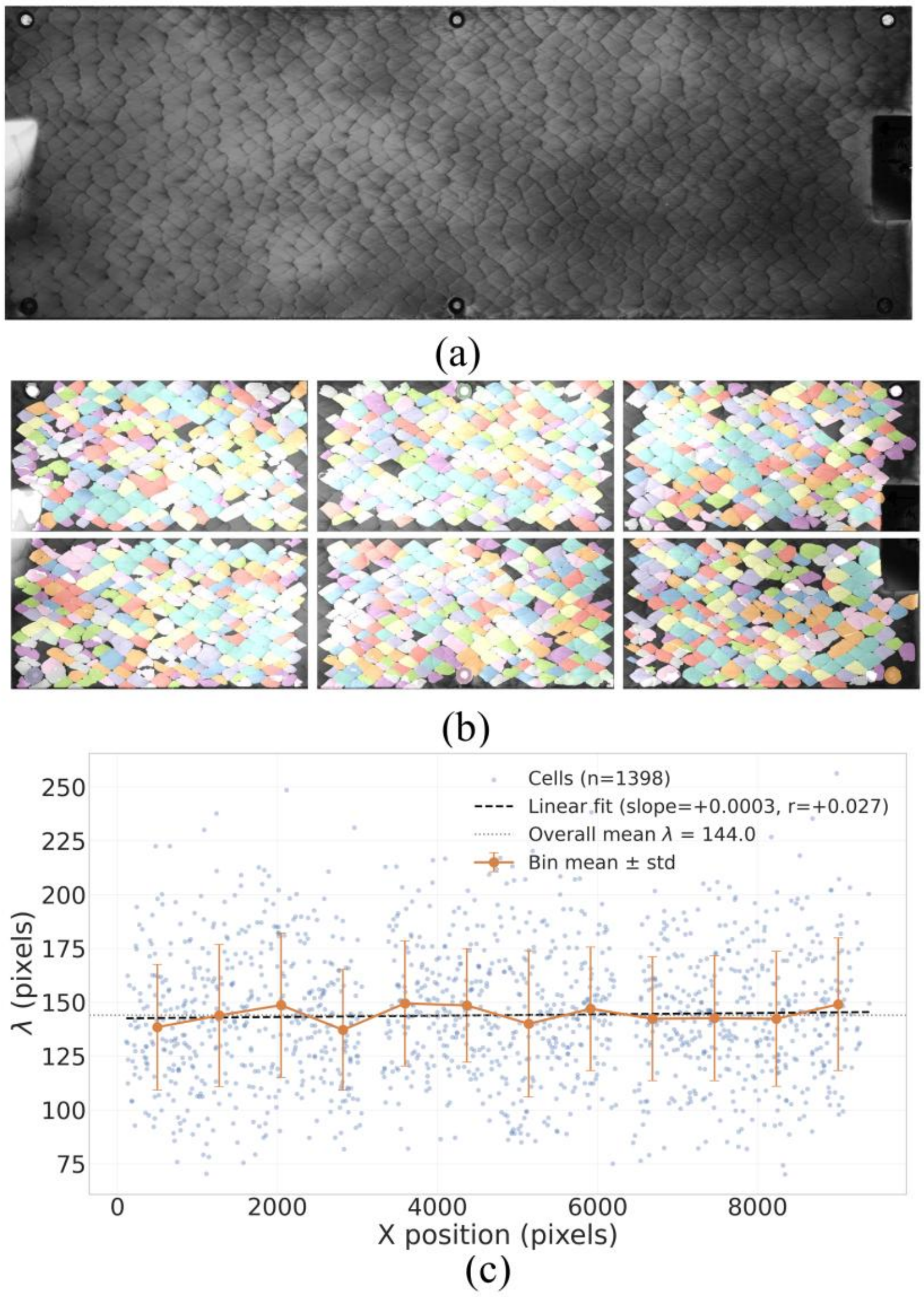

Figure 14 Irregular experimental cells: (a) original image; (b) recognition results; (c) variation of λ along the propagation direction of the detonation wave.

### 3.3.2 Quantitative Extraction of High-Order Regularity Features of Detonation Cells

In addition to size, cell regularity is a crucial metric for characterizing the effective activation energy and hydrodynamic instability (stability parameter) of combustible mixtures. To overcome the purely subjective qualitative descriptions of "regular" and "irregular" structures used in the past, recent studies have proposed various high-order quantitative indicators. To validate the proposed model's capability in extracting high-order geometric features, this section introduces the Irregularity Index (RI) proposed by Shi et al.[41] and the standard deviation of cell deflection angles($\sigma_\theta$) proposed by Genter et al.[42] The RI is typically defined based on the standard deviation of the nearest-neighbor distances within the spatial distribution of triple-point collisions, with a larger RI value indicating a more chaotic cellular network. The deflection angle $\sigma_\theta$, on the other hand, is defined as the standard deviation of the angles between the geometric major axis of each cell and the global propagation direction of the detonation wave. A smaller $\sigma_\theta$ indicates a more orderly cell arrangement, whereas a larger deflection angle implies that the propagation trajectories of the internal Mach stems are highly chaotic.

Benefiting from the minimum bounding boxes and mask contours output by the instance segmentation model, the algorithm can directly compute the major-axis deflection angle (θ) and the spatial topological network for every independent cell, thereby automatically yielding the corresponding RI and $\sigma_\theta$.Figure 15 illustrates a comparison between regular and irregular cells under varying conditions, while Table 1 quantitatively summarizes the calculated high-order metrics for the regions depicted in Figure 15.

As can be seen from the data in Table 1, for the highly regular simulated condition shown in Figure 15(a), the model-calculated RI is merely 0.06, and $\sigma_\theta$ is only 0.63°, indicating an almost perfect horizontal fish-scale-like arrangement of all cells. Conversely, in the irregular simulated condition shown in Figure 15(c), the RI increases to 0.18, and $\sigma_\theta$ surges to 10.21°, thereby accurately quantifying the geometric distortion induced by instability. Similarly, regarding the real experimental data, the RI (0.17) and $\sigma_\theta$ (3.61°) for the regular condition in Figure 15(e) are significantly lower than those for the irregular condition in Figure 15(g) ( RI=0.26, $\sigma_\theta$=6.83°).

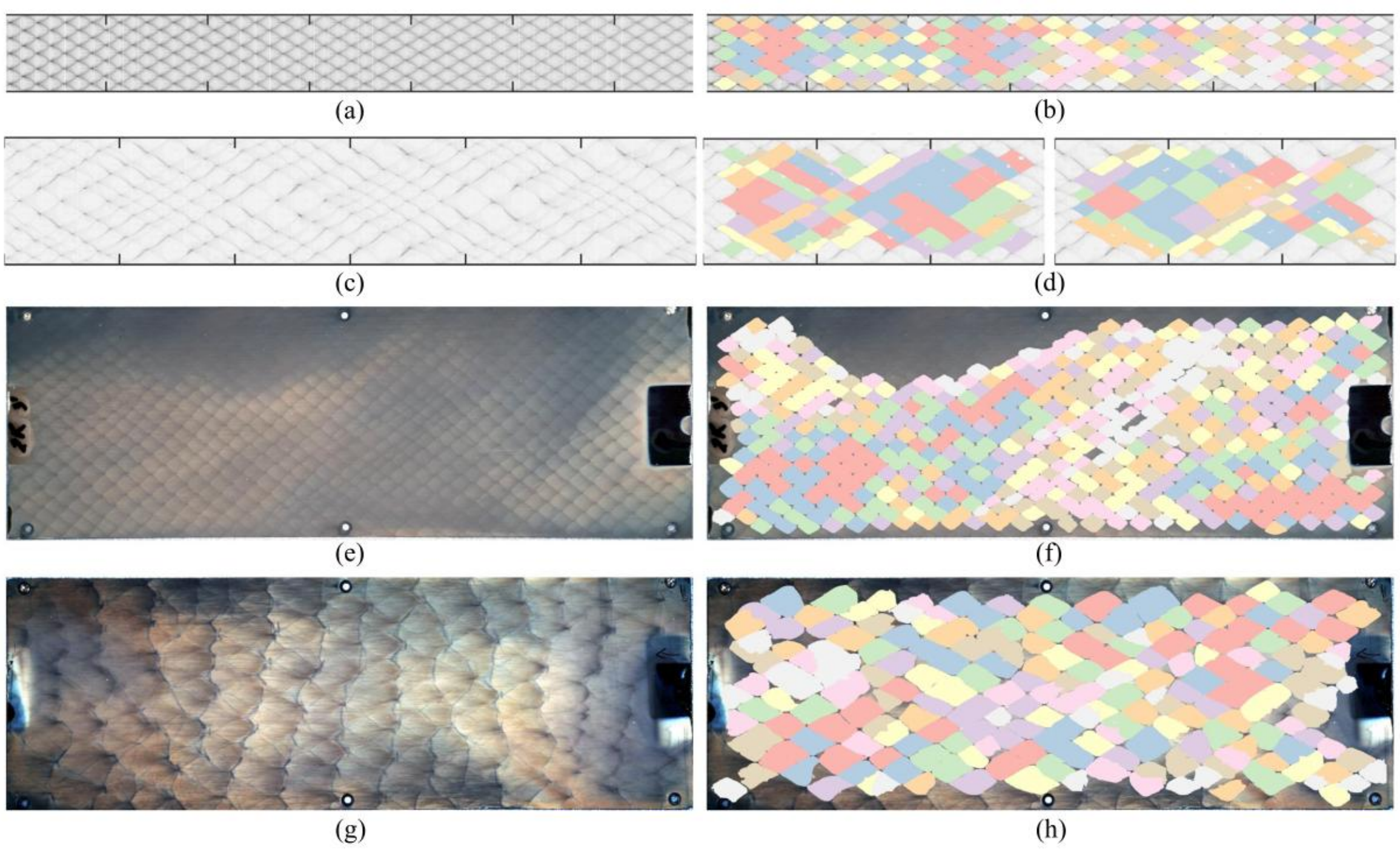

Figure 15 Visualization of original images and recognition results for regular and irregular cells under simulated and experimental conditions.

Table 1 Irregularity index (RI) and standard deviation of cell deflection angles ($\sigma_\theta$) calculated by the model for the different images in Figure 15.

| Picture | Figure 15(a) | Figure 15(c) | Figure 15(e) | Figure 15(g) |
|---|---|---|---|---|
| RI | 0.06 | 0.18 | 0.17 | 0.26 |
| $\sigma_\theta$ | 0.63 | 10.21 | 3.61 | 6.83 |

It is worth emphasizing that the data trends of the high-order features extracted by the proposed model (i.e., irregular flow fields corresponding to elevated RI and $\sigma_\theta$) are highly consistent with theoretical expectations from the physics literature[41,42]. This demonstrates that the deep learning method developed in this study is not confined merely to basic feature segmentation; rather, it serves as a powerful, objective, and fully automated quantitative diagnostic tool for fundamental physical research on detonation flow fields (e.g., correlation analyses between chemical kinetic sensitivity and three-dimensional wave system instabilities).

# 4 Conclusion

To address the challenge of quantitatively analyzing the characteristic dimensions of detonation cells, this paper developed a model for cell recognition and high-dimensional feature extraction based on a deep learning instance segmentation architecture (Mask R-CNN). Through systematic benchmark validations, evaluations on experimental data, and ablation analyses, the following principal conclusions are drawn:

(1) Accurate Pixel-Level Segmentation and Strong Robustness Against Noise: Compared to conventional image processing or semantic segmentation methods, the proposed model effectively overcomes the interferences caused by uneven soot coatings, contrast degradation, and severe cell overlapping in experimental soot foil images. The model achieves high-precision pixel-level mask prediction for each individual cell instance, yielding a global statistical mean that is highly consistent with manual reference values (relative error< 3.5%). This not only circumvents the tediousness of traditional manual measurements but also eliminates the subjective bias inherent in manual annotations, thereby providing a more scientific and objective statistical probability density distribution.

(2) Identification of Model Sensitivity Boundaries and a Standardized Processing Paradigm: Sensitivity analyses demonstrate that the model possesses excellent scale invariance, maintaining the relative error at approximately 5% even when the image resolution is downsampled to 30%. Furthermore, this study elucidates the mechanisms underlying the prior distribution shift and localized missed detections induced by an excessively high per-image object count. Consequently, a rational image patching strategy (e.g., 2 × 3 partitioning) was established. This strategy effectively circumvents the contraction of underlying feature scales within the algorithm, thereby laying the foundation for a standardized preprocessing paradigm for the future large-scale processing of complex detonation flowfield images.

(3) Realization of High-Order Data Extraction for Fundamental Detonation Research: The proposed model transcends the traditional application level of merely calculating the average cell size λ. By fully leveraging the spatial coordinates and geometric contours output by the instance segmentation, it achieves the automated tracking of the spatial evolution of cell sizes along the propagation direction of the detonation wave, thereby effectively determining the transient initiation or stable Chapman-Jouguet (CJ) propagation states of the detonation wave. Furthermore, the model enables the fully automated extraction of high-order morphological features, such as the irregularity index (RI) and the standard deviation of cell major-axis deflection angles ($\sigma_\theta$). These quantitatively extracted results align well with cutting-edge theories regarding wave system instability.

The deep learning instance segmentation-based cell recognition model developed in this study provides a practical technical solution for the automated and objective processing of experimental detonation data. This method not only effectively overcomes the inherent drawbacks of traditional manual measurements—which are highly susceptible to subjective experience and local noise interference—but also directly transforms the complex physical morphology of the cells into structured data encompassing spatial coordinates, local dimensions and deflection angles. This fully automated feature extraction capability empowers researchers to quantitatively track the transient evolution of the detonation wave along its propagation direction and to efficiently acquire high-order geometric parameters for characterizing hydrodynamic instabilities. Overall, this study not only successfully resolves the bottleneck of high-precision statistical analysis of cellular structures but also provides a robust and reliable quantitative analysis tool for future investigations into the intrinsic correlations between the chemical kinetic properties of combustible mixtures and macroscopic detonation propagation mechanisms.

# Acknowledgments

The authors would like to thank the support from the National Natural Science Foundation of China (No. 12441201). The authors are grateful to CNPq (Conselho Nacional de Desenvolvimento Científico e Tecnológico) for supporting this work through Projects 442302/2023-1 and 308915/2022-4, and to FAPERGS (Fundação de Amparo à Pesquisa do Estado do Rio Grande do Sul) for supporting this work through Project 24/2551-0001246-0.

# Appendix

To further demonstrate the generalization capability and engineering applicability of the proposed instance segmentation model, this appendix provides supplementary visualization results of cell recognition under diverse experimental conditions. Appendix Figure 1 through Appendix Figure 4 present the original soot foil records (a) and their corresponding pixel-level mask prediction results (b), respectively. These supplementary examples encompass an exceptionally broad range of cell morphological features, spanning from highly regular uniform grids (Appendix Figure 1) to severely distorted irregular structures with massive scale variations (Appendix Figure 4). Furthermore, they incorporate various background noises resulting from uneven soot coatings and variations in illumination. It is particularly noteworthy that, as illustrated in Appendix Figure 3, when the detonation flow field exhibits distinct physical boundaries (such as regions of shock diffraction or localized failure), the model demonstrates outstanding background rejection capabilities. Specifically, it accurately halts recognition at the genuine cell margins, avoiding the generation of false positive detections in blank regions devoid of trajectories. Taken together, these supplementary results intuitively reaffirm the high degree of robustness and reliability of the proposed deep learning model when processing complex and highly variable real-world experimental detonation images.

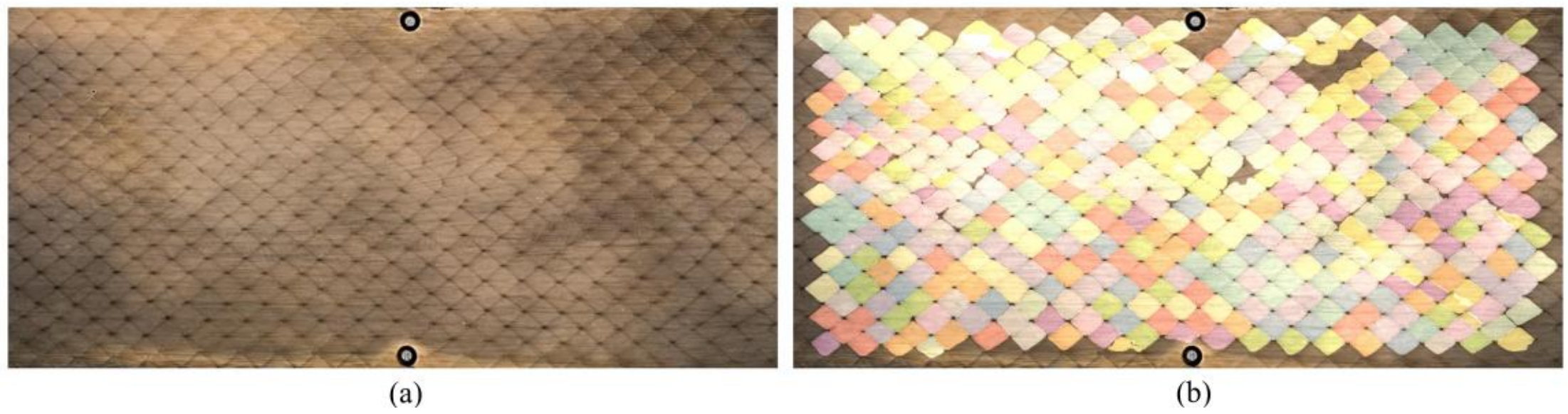

Appendix Figure 1 (a) original image; (b) visualization of recognition results

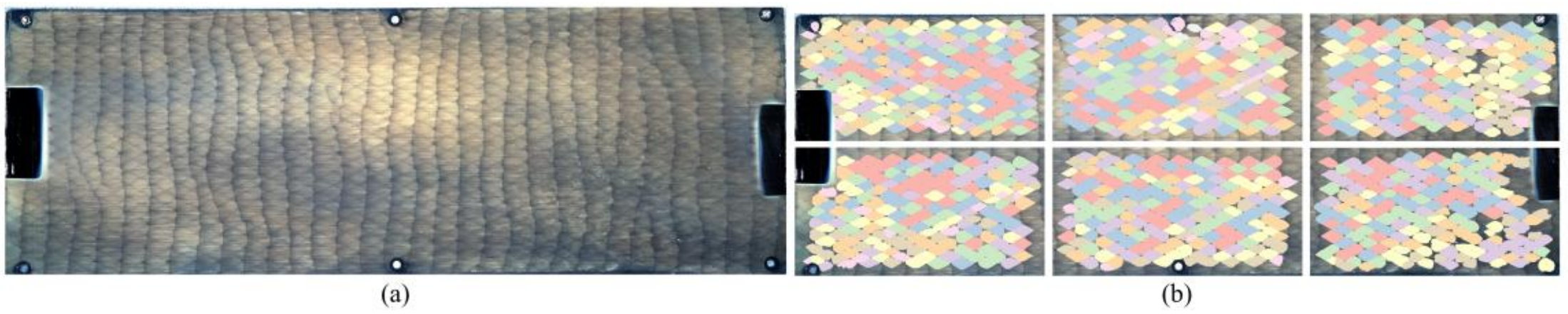

Appendix Figure 2 (a) original image; (b) visualization of recognition results

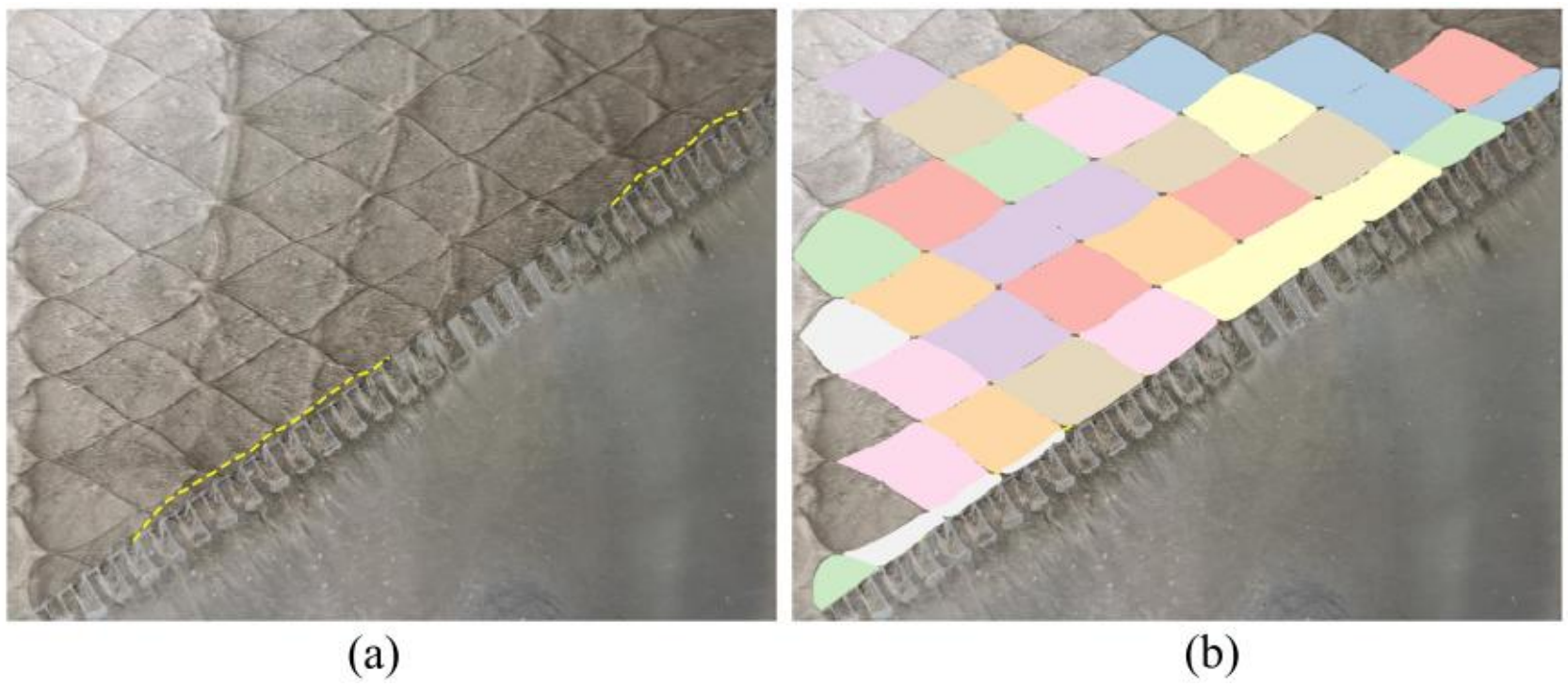

Appendix Figure 3 (a) original image; (b) visualization of recognition results

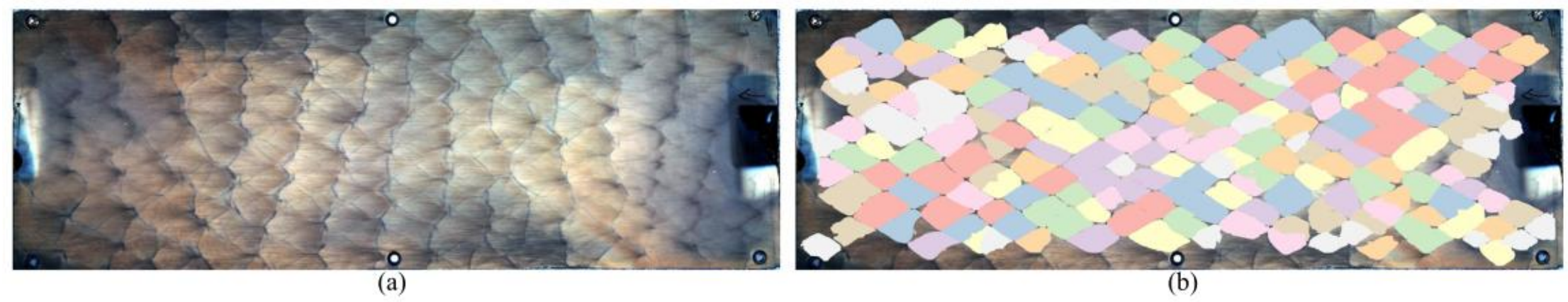

Appendix Figure 4 (a) original image; (b) visualization of recognition results